\documentclass[preprintnumbers,aps,prb,preprint]{revtex4}

\usepackage{graphicx}
\usepackage{amsmath}
\usepackage{amssymb}

\begin{document}

\title{Dissipation induced by attractive interaction in dynamic force microscopy~: contribution of adsorbed water layers.}
\author{L. Nony$^{1,^\ast}$, T. Cohen-Bouhacina$^2$, J.-P. Aim\'{e}$^2$\\
 \small{$^1$ L2MP, UMR CNRS 6137,
 Universit\'{e} d'Aix-Marseille III \\
 Facult\'{e} des Sciences de Saint-J\'{e}r\^{o}me, 13397 Marseille Cedex 20, FRANCE}\\
 \small{$^2$ CPMOH, UMR CNRS 5798, Universit\'{e} Bordeaux I\\
 351, cours de la Lib\'{e}ration, 33405 Talence Cedex, FRANCE}\\
 \small{$^\ast$ To whom correspondence should be addressed; E-mail:
 laurent.nony@l2mp.fr}\\
 \textbf{published in Surface Science 499, pp152-160 (2002)}}

\begin{abstract}
At room temperature and under ambient conditions, due to the adsorption, a water film is always present on silica surfaces. If the surface is
investigated with a scanning probe method in Contact mode, this causes the formation of a meniscus between the tip and the surface. This liquid
neck generates additional capillary forces between the nano-tip and the surface. In dynamic mode, due to the action of the oscillating tip on the
surface, the mechanical response of the adsorbed water layers can induce additional dissipation that is probed through the phase variations of the
oscillator. In the present work, we analyze by dynamic force microscopy the growth of a water film on a silica surface as a function of time. The
silica sample is first cleaned and heated at $420^\circ$C, then is exposed to dry conditions. The influence of the water film is checked with the
dynamic mode by using intermittent contact and noncontact situations. To describe the experimental observations, additional dissipation is taken
into account when the tip approaches the surface. The results of the fits allow the evaluation of the dissipation induced by the attractive
interaction between the tip and the silica surface related to the adsorption of water molecules on surface as a function of time. Results are
compared to previous tribological studies performed in Contact mode and infra-red spectroscopy measurements on the silica for which the key
parameter was the surface temperature instead of time. The two experimental results are in good agreement.

\textbf{keywords :}\ Atomic Force Microscopy, Adsorption kinetics, Growth, Silicon oxides.\\ PACS 05.45.-a, 07.79.Lh, 45.20.Jj
\end{abstract}

\maketitle

\section{Introduction}

During the last decade, Dynamic Force Microscopy (DFM) used in the Tapping
mode has been found as a suitable tool to investigate surfaces morphology
and mechanical properties of soft materials. The technics has been widely
used to investigate a wide range of samples including polymers \cite
{Leclere96,Magonov97}, biological materials \cite{Rivetti96,Shlyakhtenko99}
or organic layers \cite{Barrat92,Vallant98}. Taking advantage of the
sensitivity of the oscillating tip-cantilever system (OTCL) at the proximity
of the surface, various physical properties can be investigated including
mechanical properties \cite{Tamayo97,Haugstadt98,Aime1_99,Marsaudon00},
adhesion \cite{Dubourg00}, forces mapping \cite{Garcia99,Nony99,Nony01} and
dissipation processes \cite
{Tamayo98,Gotsmann99,Gauthier99,Dorofeyev99,Bennewitz00,Aime01}. As a
complementary approach, experiments performed in Contact mode were widely
used to probe chemical treatments of surfaces and evolution of surface
properties. Therefore, numerous works were dedicated to investigate adhesion
and friction variations as a function of changes of the surface properties
under various conditions~: humidity rate, vacuum or controlled atmosphere
\cite{Binggeli95,Weisenhorn89,Scherge98}.

Because DFM is a very sensitive tool probing forces is noncontact and
intermittent contact situations, it is a suitable approach to investigate
local properties without inducing severe damage on the surface. In
particular, liquid layers and the influence of capillary forces have been
extensively investigated \cite{DePablo00,Luna00}. The presence of a liquid
layer from molecular thickness up to a few nanometers can significantly
modify the behavior of the OTCL. In ref.\cite{Luna00}, M.Luna et al. discuss
the influence of the conservative and non conservative forces due to the
water adsorption onto various surfaces on the variation of the resonance
frequency and quality factor of the OTCL.

In the present work, we use DFM to investigate the growth of a water film on
a silica surface at room temperature as a function of the exposure time. To
do so, we use recent theoretical developments describing the influence of
the noncontact dissipation (NC dissipation) on the observed phase delay
during a Tapping experiment \cite
{Aime01,Couturier01,Couturier2_01,Couturier3_01,Boisgard01}. The silica
surface is first cleaned and heated under oxygen flow at $420~%
{{}^\circ}%
$C during $90$ minutes. Then, it is put into a glove box under controlled
atmosphere assuming that the water covering goes along. Evolution of the
surface is measured by recording approach-retract curves versus time that
give the variations of the oscillation amplitude and phase as a function of
the tip-surface distance.

\bigskip

In the first part of the paper, we briefly recall the model used to take
into account the surface modifications and NC dissipation processes. The
second part is devoted to experimental results obtained with the cleaned and
heated silica surface and a comparison with the theoretical developments.
The experimental results are compared to those obtained with a previous
tribological study performed in Contact mode as a function of the sample
temperature. The last part of the paper is a discussion of the results
obtained.

\section{Modeling the OTCL's behavior}

The present section gives a synthesis of the theoretical developments
allowing analytical expressions of the nonlinear equations of motion of the
OTCL --amplitude and phase variations as a function of the distance between
the tip and the surface-- to be obtained in order to fit the experimental
data. The aim of these equations is to reproduce the observed
approach-retract curves and more particularly the increase of the phase
delay due to the attractive interaction between the tip and the surface. By
using a sphere-plane geometry with an attractive Van der Waals interaction
\cite{Israelachvili92} describing the interaction between the tip and the
surface~:

\begin{equation}
V\left[ z\left( t\right) \right] =-\dfrac{HR}{6\left[ D-z\left( t\right) %
\right] }\text{,}  \label{equpotVanderWaals}
\end{equation}
the equations of motion are given by \cite{Nony99}~:

\begin{equation}
\left\{
\begin{array}{l}
\cos \left( \varphi \right) =Qa(1-u^{2})-\dfrac{aQ\kappa _{a}}{3\left(
d^{2}-a^{2}\right) ^{3/2}} \\
\sin \left( \varphi \right) =-ua
\end{array}
\right.  \label{equcossin}
\end{equation}
In equ.\ref{equpotVanderWaals},$\ D$ is the distance between the sample and
the equilibrium position at rest of the OTCL and $z(t)=A\cos \left( \omega
t+\varphi \right) $ the position of the tip at time $t$ with the driven
frequency $\omega $. $H$ and $R$ are the surface Hamaker constant and the
tip's apex radius, respectively. From equ.\ref{equcossin}, one can extract
the relationship between the oscillation amplitude and the tip-surface
distance~:

\begin{equation}
d_{_{\pm }}=\sqrt{a^{2}+\left[ \frac{Q\kappa _{a}}{3\left\{ Q\left(
1-u^{2}\right) \mp \sqrt{1/a^{2}-u^{2}}\right\} }\right] ^{2/3}}
\label{equdattr}
\end{equation}
In equ.\ref{equdattr},$\ a=A/A_{0}$, $d=D/A_{0}$ and $u=\omega /\omega _{0}$
are the reduced amplitude, distance and frequency, respectively. $A_{0}$ and
$\omega _{0}$ are the resonance amplitude and resonance frequency of the
OTCL far from the surface, respectively. $Q$ is the quality factor of the
OTCL given by $Q=\omega _{0}/\beta _{0}$. $\beta _{0}$ is the OTCL damping
coefficient in air. $\kappa _{a}=HR/\left( k_{c}A_{0}^{3}\right) $ is a
dimensionless parameter which is characteristic of the nonlinear coupling
with $k_{c}$, the cantilever' stiffness. Thus, varying $\kappa _{a}$ with $%
A_{0}$ is equivalent to vary the strength of the attractive interaction \cite
{Nony99}.

\begin{figure}[h]
\includegraphics[width=10cm]{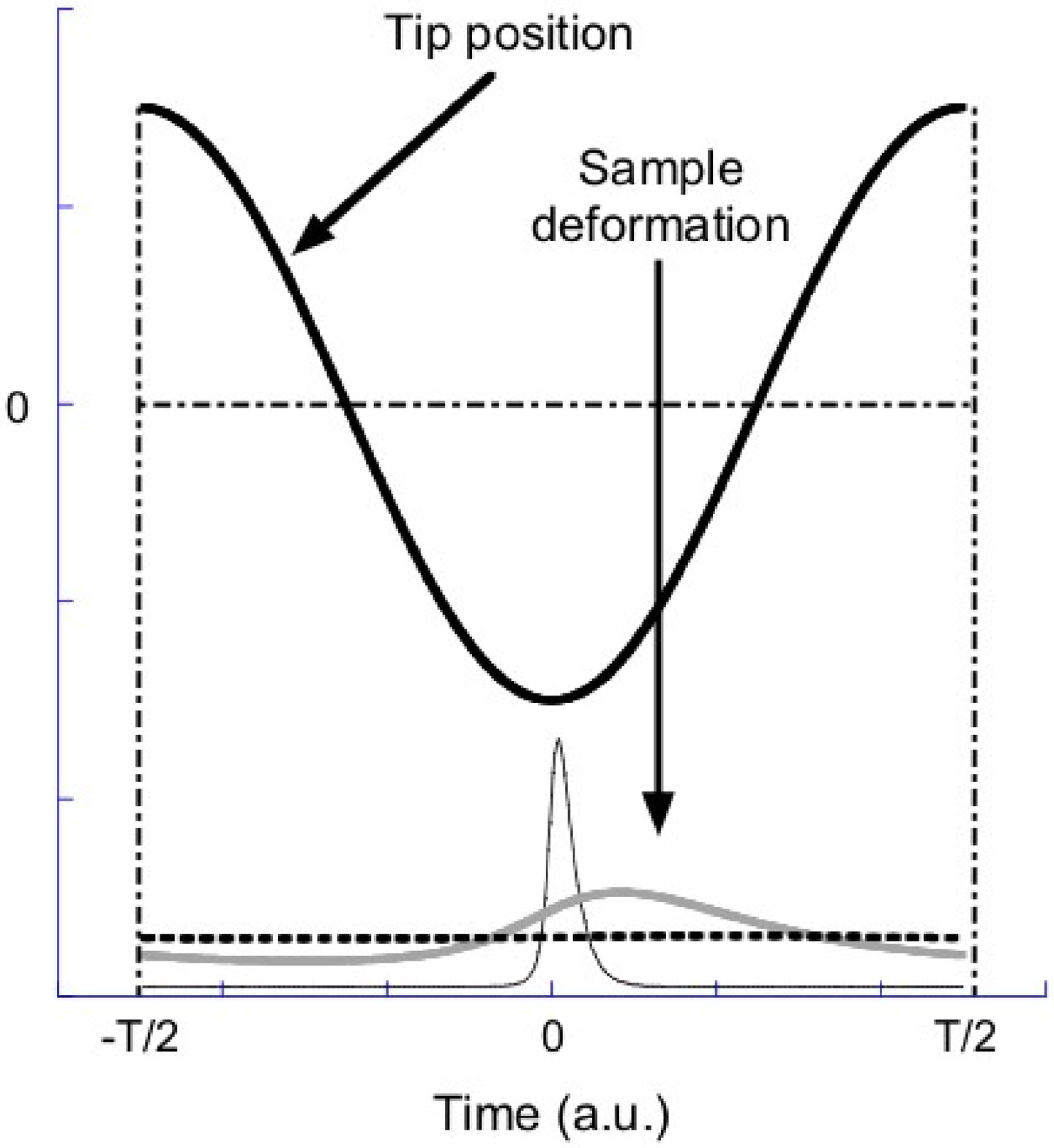}\\
  \caption{Sketch of the action of the
oscillating tip on the sample. The viscoelastic behavior of the surface ($%
\protect\gamma _{s}$, $k_{s}$) implying that it can be deformed under the
action of the tip-surface attractive force. Depending on the value of its
relaxation time $\protect\tau _{s}=\protect\gamma _{s}/k_{s}$ with regard to
the period $T$ and the residence time $\protect\tau _{res}$ (long, $\protect%
\tau _{s}\gg T$~: thick black dashed line, short, $\protect\tau _{s}\ll \protect\tau _{res}$~: thin black line or intermediary~: thick grey line),
various expressions of the additional dissipation $\protect\beta _{int}$ can be obtained (see text).}\label{Figtoustemps}
\end{figure}

In order to fit the experimental data, e.g. amplitude and phase variations
as a function of the distance between the tip and the surface, it was shown
that the introduction of an additional dissipation when the tip does not
touch the surface was a necessary requirement that explained the increase of
the phase delay \cite{Aime01,Couturier2_01}. The additional dissipation is
modeled as being the consequence of the OTCL energy loss due to the
mechanical response of the substrate induced by the attractive interaction
with the tip \cite{Aime01,Couturier01,Couturier2_01,Couturier3_01,Boisgard01}%
. The substrate is modeled as a viscoelastic, thus described by its local
stiffness $k_{s}$ and damping coefficient $\gamma _{s}$ with a relaxation
time $\tau _{s}=\gamma _{s}/k_{s}$. These works give a detailed explanation
of the origin of an additional damping coefficient $\beta _{int}$. Depending
on the value of $\tau _{s}$ with regard to the characteristic times of the
OTCL --its oscillation period $T$ and the residence time of the tip at the
proximity of the surface, $\tau _{res}$-- two asymptotic regimes of
dissipation can be deduced. $\tau _{res}$ is the residence time of the tip
in the vicinity of the surface, e.g. when the coupling between the OTCL and
the surface is large and is given by $\tau _{res}\simeq \dfrac{T}{\pi }\sqrt{%
2\Delta /A}$ \cite{Aime01,Couturier2_01}, where $\Delta \simeq D-A$.
Depending on the values of $\tau _{s}$ (see fig.\ref{Figtoustemps}), two
asymptotic behaviors are calculated. Results are \cite{Boisgard01}~:

\begin{eqnarray}
\text{Short relaxation times }\tau _{s} &\ll &\tau _{res}\text{ : }\beta
_{int}\left( \Delta ,A\right) \simeq \frac{\tau _{s}K^{2}}{k_{s}}\frac{%
\omega _{0}^{2}}{12\sqrt{2}k_{c}}\times \frac{1}{\Delta ^{9/2}A^{3/2}}
\label{equtempscourts} \\
\text{Long relaxation times }\tau _{s} &\gg &T\text{ : }\beta _{int}\left(
\Delta ,A\right) \simeq \frac{K^{2}}{\gamma _{s}}\frac{1}{2k_{c}}\times
\frac{1}{\Delta ^{7/2}A^{5/2}}  \label{equtempslongs}
\end{eqnarray}
with $K=HR/6$. Thus the NC dissipation processes are controlled by the
strength of the attractive interaction through a square dependence with the
term $\left( HR\right) ^{2}$. The limit of short relaxation times is $\tau
_{s}\rightarrow 0$ for which the sample behavior is elastic-like. In that
case, the local deformation of the surface follows the action of the
oscillating tip without any phase delay (see fig.\ref{Figtoustemps}). As a
consequence, the dissipated energy goes to zero. For long relaxation times,
the mechanical susceptibility of the sample scales as $1/\gamma _{s}$, then
also the additional dissipated energy.

The total damping coefficient of the OTCL can be written as an equivalent
damping term \cite{Aime01,Boisgard01}~:

\begin{equation}
\beta _{eq}(\Delta ,A)=\beta _{0}+\beta _{int}(\Delta ,A)\text{,}
\end{equation}
leading to the variations of the sine of the phase (see equ.\ref{equcossin})
on the form~:

\begin{equation}
\sin \varphi =-ua\left( \dfrac{\beta _{eq}(\Delta ,A)}{\beta _{0}}\right)
\label{equsinfidissipNC}
\end{equation}
Equ.\ref{equsinfidissipNC} is the one that is used to fit the experimental
phase variations.

\bigskip

The variation of the amplitude without including an additional dissipation
is straightforwardly obtained from equ.\ref{equdattr}. In accordance with
the known result that the dissipation reduces the nonlinear effects \cite
{Thompson91,Aime3_99}, a first consequence is a drastic reduction of the
size of the hysteresis loop \cite{Aime01}. When the OTCL approaches the
surface, the amplitude variations are similar. The main difference appears
at the bifurcation spot where the magnitude of the amplitude jump is
slightly frustrated. We have shown that equ.\ref{equdattr} was of some help
to estimate the value of the $HR$ product \cite{Aime01}.

\section{Results}

\subsection{Experimental conditions}

\textbf{Dynamical mode~: }The silica substrate is cleaned as detailed
elsewhere \cite{sandrine96} and heated under oxygen flow at $420\mathbf{~}%
{{}^\circ}%
$C during $90$ minutes\ in order to remove the water layers and
contaminants. Then the sample is put into a glove box under nitrogen
controlled atmosphere with one ppm (part per million) of water and oxygen
and at room temperature. Experiments of approach-retract curves were
performed with a Nanoscope III \cite{Digital} operating in Tapping mode into
the glove box. In such conditions, the cantilever behavior is very stable. A
commercial silicon tip-cantilever TESP-NCL-W from Nanosensors \cite
{Nanosensors} is used, with an announced stiffness of about $40$~N.m$^{-1}$,
a measured resonance frequency of $185130$~Hz and a quality factor of $500$.
Particular attention was focussed on the quality of the OTCL whose behavior
has to be harmonic for quantitative measurements. Technical remarks
concerning the determination of the cantilever parameters and data
treatments are detailed in ref.\cite{Nony99}. The experiments were performed
at $184944$~Hz, corresponding to a phase $\varphi _{free}\simeq -45%
{{}^\circ}%
$ and an oscillation amplitude $A_{free}=A_{0}/\sqrt{2}\simeq 0.707A_{0}$
for large tip-surface distances. The subscript ``free'' means that the
oscillation conditions are measured at a tip-surface distance for which the
tip does not interact with the surface, typically distances of ten
nanometers or more.

To investigate the influence of water layers, all the experiments presented
in this paper were performed with the same cantilever over three days. Over
those three days, no noticeable variation of the OTCL parameters (resonance
frequency and quality factor) was observed.

\bigskip

\textbf{Contact mode~:} Nanotribological measurements are detailed in a
previous paper \cite{Bouhacina00}. Here, we briefly recall the experimental
conditions and present a synthetic result which will then be compared to DFM
results (see discussion). The silica substrate is also cleaned and heated at
$420~%
{{}^\circ}%
$C under oxygen flow, then exposed to dry conditions. All experiments were
performed in a glove box under dry conditions with nitrogen flow. This
condition is necessary since experiments performed in air (under
uncontrolled humidity rate) were not as much reproducible to extract a
significant variation of the adhesion and friction as a function of the
temperature. Force curves and friction loops were recorded at temperature
varying between $25~%
{{}^\circ}%
$C and $170~%
{{}^\circ}%
$C \cite{Bouhacina00}.

\subsection{Results obtained in Tapping mode}

\subsubsection{Noncontact situations~: evaluation of the $HR$ product}

When the sample is taken out of the furnace, corresponding to the reference
time --set time zero--, approach-retract curves are recorded every two or
three hours, three days long. During the approach-retract curve, when the
bistable behavior occurs, the tip is at a vertical location far enough from
the surface, in the range of one nanometer or more, to reduce the
contribution of the NC dissipation. As a consequence, one can estimate that
the bifurcation is only controlled by $HR/k_{c}$ and $A_{0}$ through the $%
\kappa _{a}$ dependence. Nevertheless, note that the strength of the
attractive force scaling as $\sqrt{1/A}$ \cite{Aime1_99} (see also the $A$
dependence in $\tau _{res}$), a too low value of the amplitude can not be
used to perform the fits. Therefore, the curves for which the noncontact
(NC) situations occur allow to get an evaluation of the $HR/k_{c}$ product
from equ.\ref{equdattr}. For this set of experiments, the amplitude at which
the first NC situation happens is $A_{free}=19$~nm $\left( A_{0}=27~\text{nm}%
\right) $. To simplify the analysis and use the fit of the experimental NC
curves, a typical value of the amplitude used is $A_{free}=12.5$~nm $\left(
A_{0}=18~\text{nm}\right) $. The experimental measurements are then compared
at different times.

\begin{figure}[h]
\includegraphics[width=10cm]{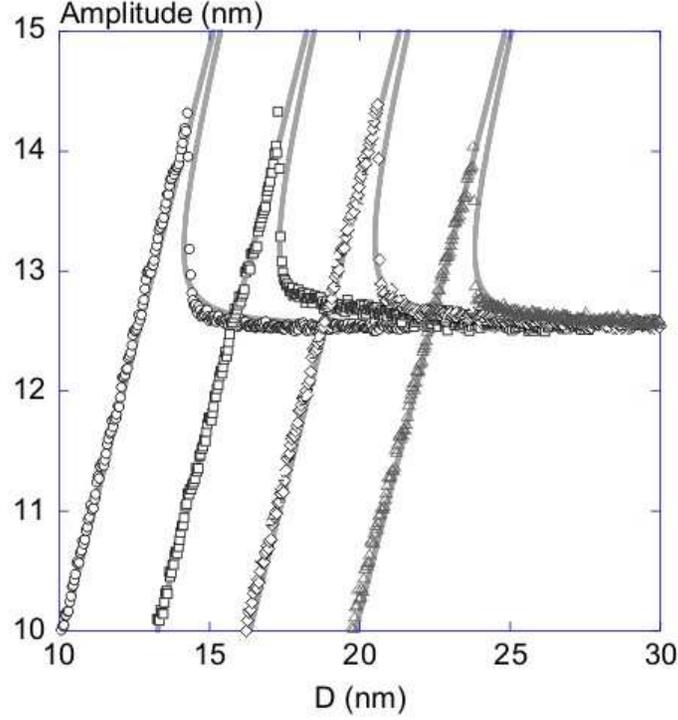}\\
  \caption{Experimental noncontact amplitude
variations (open black symbols) versus tip-surface distance for $%
A_{free}=12.5~$nm ($A_{0}=18$~nm) as a function of time and their
comparisons with theoretical curves deduced from equ.$3$ (continuous grey
lines). The theoretical curves are calculated in order to fit the location
of the bifurcation, thus giving an evaluation of $HR/k_{c}$. The general
experimental parameters are given in the text. Circles, $time=0.7$~hours;
squares, $time=7$~hours; diamonds, $time=30$~hours; triangles, $time=76$%
~hours. The horizontal locations of the curves have been arbitrarily shifted in order to show all the variations.}\label{figfit_expenc}
\end{figure}

In fig.\ref{figfit_expenc} are reported comparisons between experimental and
theoretical curves for different amplitudes and different times. The equ.\ref
{equdattr} used to fit the experimental data gives a reasonable agreement.
The average value of the coefficient $HR/k_{c}$ extracted from the fits is $%
0.46.10^{-27}$~m$^{3}$. Thus, by taking $k_{c}\simeq 40$~N.m$^{-1}$, we get $%
HR\simeq 18.10^{-27}$~J.m. Taking into account the fact that the very first
NC situation is obtained for $A_{free}=19$~nm, it's expected that the radius
of the tip is not very small since we usually observe NC situations on the
silica for $A_{free}\simeq 10$~nm, or less \cite{Nony99}. Thus in that case,
let's consider $R\simeq 20$~nm. As a consequence, we get for the silica $%
H\simeq 10^{-19}$~J. This evaluation is in agreement with a previous result
(see below).

\begin{figure}[h]
\includegraphics[width=10cm]{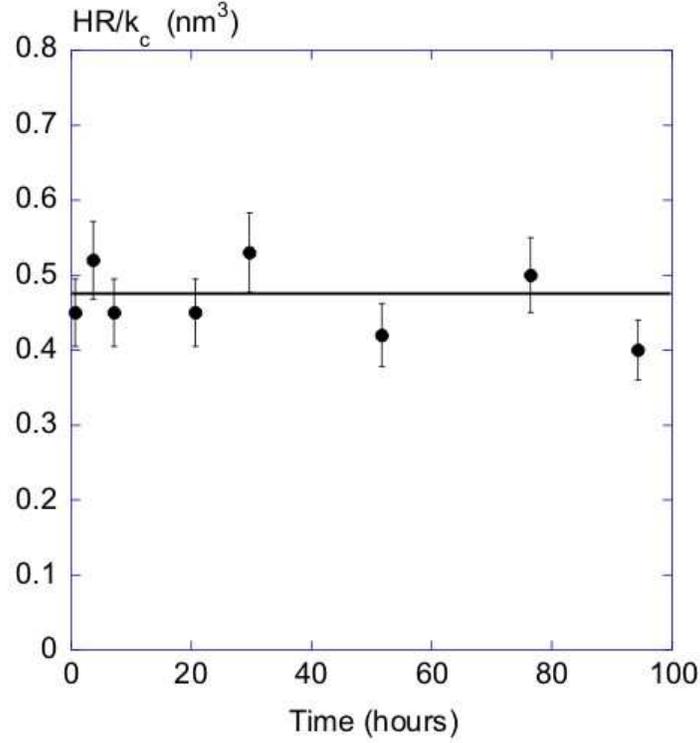}\\
  \caption{Values of $HR/k_{c}$ deduced from the fits of the experimental amplitudes variations as a function of
time. The fits were performed on the curves obtained for $A_{free}=12.5~$nm (see
fig.$2$). $HR/k_{c}$ doesn't evolve with time and the average value is $%
0.46.10^{-27}~$m$^{3}$. Since $R$ and $k_{c}$ are constant parameters, a constant value of $HR/k_{c}$ means that the Hamaker constant remains
nearly the same, $H\simeq 10^{-19}~$J (see text).}\label{figsynthr}
\end{figure}

The fits do not provide an accurate quantitative value of $HR$ but allow to
evaluate the variation of the strength of the attractive interaction between
the tip and the surface as a function of time. The result of the fits
obtained at different times for $A_{free}=12.5$~nm are given in fig.\ref
{figsynthr}. Any variation of the value of $HR/k_{c}$ is noticed three days
long. Since the radius of the tip and the cantilever stiffness are constant
parameters, a constant value of $HR/k_{c}$ means that the Hamaker constant
remains nearly the same. Therefore, whatever the evolution of the silica
surface during those three days, this evolution does not lead to a
noticeable variation of the attractive interaction between the tip and the
surface.

The same method was used to compare the attractive interaction between a tip
and a silica surface and the same tip and a grafted silica surface. From the
fits, it was deduced that $\left( HR\right) _{silica}\simeq 5.10^{-27}$~J.m
and $\left( HR\right) _{grafted\text{ }silica}\simeq 11.5.10^{-27}$~J.m \cite
{Aime01}. While it is difficult to evaluate the error on the $H$ values, the
use of equ.\ref{equdattr} appears quite helpful either to compare two
different surfaces or to evaluate the relative evolution of the surface
properties for a given tip.

\subsubsection{Intermittent contact situations~: evolution of the phase delay%
\label{subsubsectionIC}}

As detailed in refs.\cite{Aime01,Boisgard01}, in Tapping mode it is more
convenient to use intermittent contact (IC) situations to evaluate the
contribution of the NC dissipation on the phase delay. The main reason is
that one needs to use an average distance between the tip and the surface,$%
\Delta $ , that gives an order of magnitude of the strength of the
attractive interaction $HR/\left[ 6\left( D-z\left( t\right) \right) ^{2}%
\right] \simeq HR/\left( 6\Delta ^{2}\right) $. To simplify our evaluation,
we consider a fixed value of $\Delta $, typically $\Delta =\bar{\Delta}=0.5$%
~nm. These assumptions are easily achieved with a large oscillation
amplitude. Note that for IC situations, $\Delta =D-A$ is less than the
percent of the amplitude and therefore $\bar{\Delta}\ll A$. Thus, while for
pure NC situations one has two varying parameters, the closest distance $%
\Delta $ and the oscillation amplitude $A$, for IC situations a fixed value
of the closest distance $\bar{\Delta}$ appears as a reasonable working
assumption to evaluate the average contribution of the attractive
interaction.

\begin{figure}[h]
\includegraphics[width=10cm]{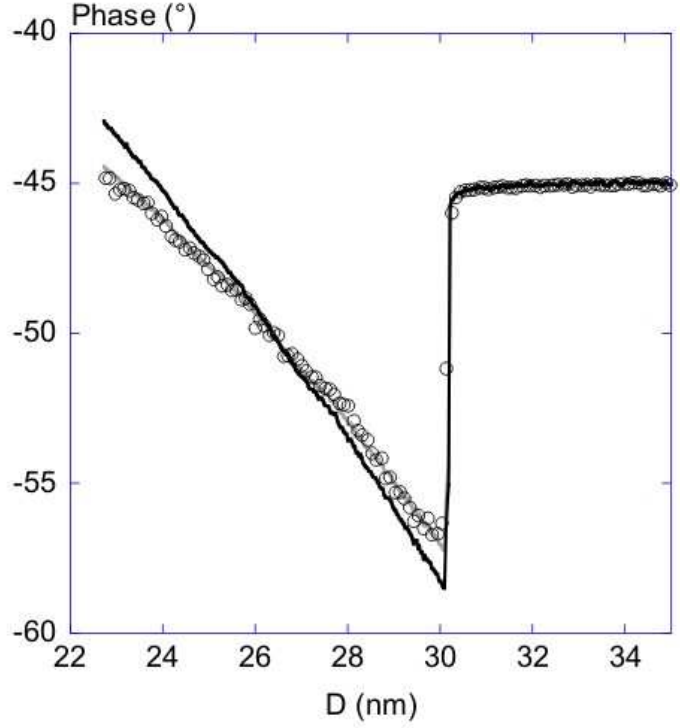}\\
  \caption{Experimental phase variation
(open black circles) versus tip-surface distance and its comparison with
theoretical curves deduced from the long relaxation times model, $A^{-5/2}$
equ.$5$ (continous grey line) and the short relaxation times model, $%
A^{-3/2} $ equ.$4$\ (continuous black line). The experimental parameters are $A_{free}=36~$nm ($A_{0}=51~$nm) and
$time=7~$hours.}\label{figpowerdea}
\end{figure}

Experimental phase variations at different times and the theoretical curves
obtained from equ.\ref{equsinfidissipNC} with the large values of $\tau _{s}$
($\beta _{int}\propto A^{-5/2}$, equ.\ref{equtempslongs}) are shown in fig.%
\ref{figfit_expeci}. The phase jump increases as a function of time. This is
predicted to be the consequence of an increase of the NC dissipation \cite
{Aime01} (see also discussion). The phase variations are fitted as a
function of the observed variations of the oscillation amplitude. There is
only one varying parameter which is $1/\gamma _{s}$, the others parameters
being evaluated ($HR$), or estimated ($k_{c}$, $\bar{\Delta}$). All the
theoretical curves are obtained with the power law $A^{-5/2}$. Fits
performed with the other asymptotic regime, short relaxation times, with a
power law $A^{-3/2}$ (equ.\ref{equtempscourts}) can be separated
unambiguously, as shown with fig.\ref{figpowerdea}.

\section{Discussion}

\begin{figure}[h]
\includegraphics[width=10cm]{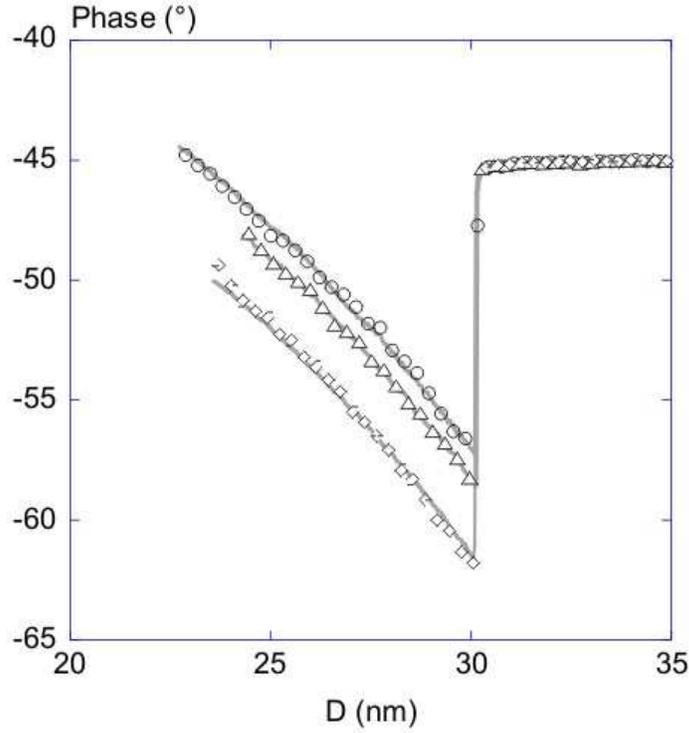}\\
  \caption{Experimental intermittent contact
phase variations (open black symbols) versus tip-surface distance as a
function of time and their comparisons with theoretical curves (continuous
grey lines) deduced from equ.$7$. The curves were obtained with $%
A_{free}=36~ $nm ($A_{0}=51~$nm). The other parameters are~: circles, $%
time=7 $~hours; triangles, $time=21$~hours; diamonds, $time=90$~hours. The increase of the phase jump as a function of time is characteristic of
an increase of the dissipated energy.}\label{figfit_expeci}
\end{figure}

\begin{figure}[h]
\includegraphics[width=10cm]{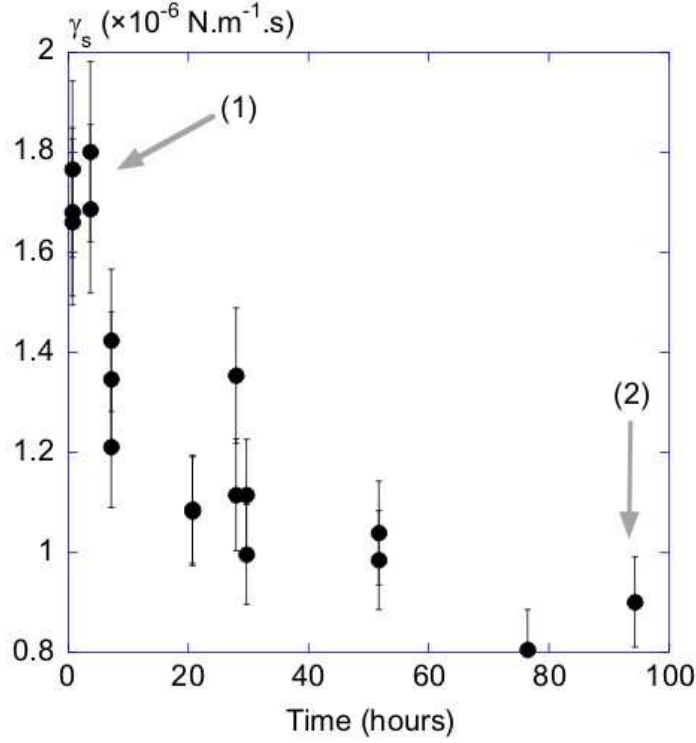}\\
  \caption{Evolution of the $\protect\gamma %
_{s}$ parameter with time deduced from the fits of the experimental phase
variations with the long relaxation times model. The values of $\protect%
\gamma _{s}$ depend on the evaluation of $HR$ and $\bar{\Delta}$ (see text). The net decrease observed, corresponding to an increase of the NC
dissipation, is the consequence of the increase of the quantity of water molecules adsorbed on the silica surface with time. The arrows 1 and 2
indicate the time-temperature equivalence that can be extracted by comparing the Tapping mode and Contact mode experiments\ (see text). For
instance, arrow 1 indicates a structure of the water layer corresponding to a temperature above $100^\circ$C, whereas arrow 2 corresponds to a
structure of the water layers observed at room temperature (see fig.$7$).}\label{figevoldissip}
\end{figure}

The mechanical susceptibility $1/\gamma _{s}$ extracted from the fits
exhibits a marked evolution versus time (fig.\ref{figevoldissip}). Since the
fits of the $HR$ product lead to a constant value (fig.\ref{figsynthr}),
this result indicates a decrease of the surface damping coefficient. It is
noteworthy recalling that the amount of NC dissipated energy corresponds to
the ability of the surface to exhibit a viscoelastic displacement induced by
the attractive interaction. Therefore for surface relaxation times larger
than that of the oscillation period, when $\gamma _{s}$ decreases, a surface
displacement proportional to $1/\gamma _{s}$ leads to an increase of the NC
dissipated energy. This is what is straightforwardly observed with the
increase of the phase jumps of the curves with time (fig.\ref{figfit_expeci}%
).

Such a variation of $\gamma _{s}$ indicates an evolution of the surface
structure and properties as a function of time. Just after the heating
process, only a few water molecules, if any, cover the surface and are
tightly bounded to the silica. In that case, the water layer might be
considered as an amorphous layer with long relaxation times corresponding to
a high damping coefficient. The mechanical susceptibility of these molecules
is weak so that the surface is weakly perturbed by the tip-surface
attractive interaction. When the sample is exposed, even in a dry
atmosphere, water molecules are more and more adsorbed on the surface,
leading to a more fluid-like behavior of the surface with a higher molecular
mobility and, in turn, a shorter relaxation time. Nevertheless, as shown
with fig.\ref{figpowerdea}, the use of the short relaxation times model
doesn't provide a good agreement with the observed phase variations. Even $%
90 $ hours after the beginning of the experiments, it's still possible to
fit the experimental data with the long time model (fig.\ref{figfit_expeci},
diamonds symbols). In ref.\cite{Aime01}, the experimental phase variations
on the silica could not be fitted with the long times model, but only with
the short times one. But in that case, no thermal treatment of the silica
had been used and the surface was placed into the glove box for many days.
From these results, one can deduce that the quantity of water layers on this
latter surface should be larger than the one on the thermally treated
surface, even $90$ hours after the beginning of the experiments.

While equs.\ref{equtempscourts} or \ref{equtempslongs} aim at describing
pure NC situations, questions rise about the origin of the information which
is measured within Tapping experiments. As explained above, in Tapping mode,
one has two varying parameters to fit the experimental variations~: $A$ and $%
\Delta $. $A$ varying during the approach-retract curves, $\Delta $ too.
This leads to complex situations if one wants to extract NC dissipation
parameters from the pure NC situations. This is mainly the reason why, IC
situations are used, $\Delta $ being fixed to an average value, $\bar{\Delta}
$, of about a few angstroms. But as a consequence, $\bar{\Delta}$ might be
as well fixed to the so-called contact distance \cite{Israelachvili92}, $%
d_{c}\simeq 0.165~$nm, thus changing the magnitude of $\gamma _{s}$. For
instance, by choosing $\bar{\Delta}=0.25$~nm instead of $\bar{\Delta}=0.5$%
~nm, the value of $\gamma _{s}$ obtained would be $11$ times larger because
of the $\bar{\Delta}^{-7/2}$ dependence in equ.\ref{equtempslongs}. This
makes absolute quantitative measurements difficult to achieve. As a
consequence, the magnitude of $\gamma _{s}$ reported in fig.\ref
{figevoldissip} is only indicative since it could has been one order of
magnitude larger.

\bigskip

\begin{figure}[h]
\includegraphics[width=10cm]{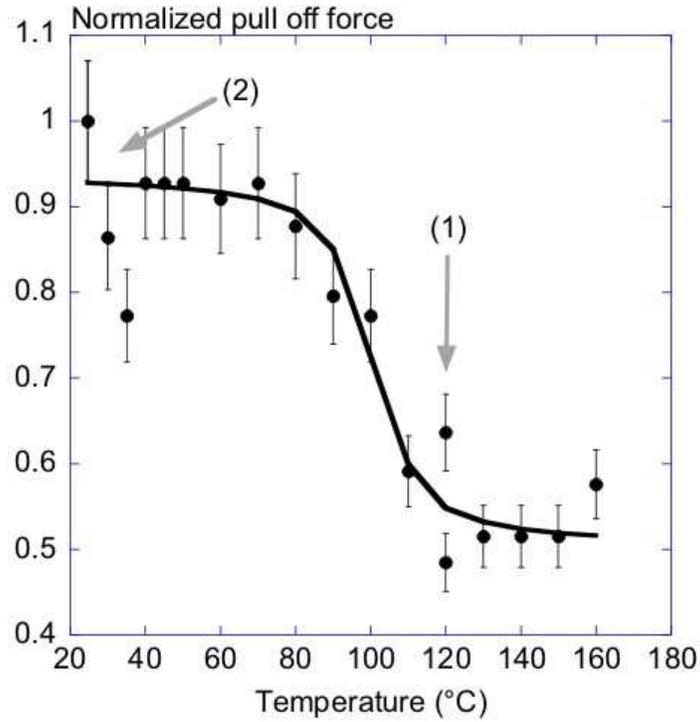}\\
  \caption{Evolution of the normalized pull
off force as a function of the silica sample temperature deduced from
Contact mode experiments \protect\cite{Bouhacina00}. The normalization pull
off force is $82~$nN. The black line is a guide line for eyes. The arrows 1
and 2 indicate the time-temperature equivalence that can be extracted by
comparing the Tapping mode and Contact mode experiments\ (see text and fig.$%
6 $).}\label{figevoladh}
\end{figure}

The DFM results are in excellent agreement with those obtained from Contact mode and FTIR spectroscopy \cite{Bouhacina00}. In these experiments,
we had focused on the influence of capillary effects due to the presence of water layers on the nanotribological properties as a function of the
sample temperature. The variations of the pull off force and of the frictions forces were reasonably well explained by assuming the creation of a
water meniscus between the tip and the surface. When the tip is pulled off the surface (adhesion force) or moved laterally (tangential force or
friction), a gap is created between the tip and the silica surface and is filled with a liquid phase. The AFM measurements show a net decrease of
the measured forces as the temperature increases. As an example, the decrease of the pull off force as a function of the temperature is shown in
fig.\ref{figevoladh}. The abrupt decrease observed around $100^\circ$C suggests that capillary forces provide a major contribution to the pull off
and tangential forces at room temperature.

The nature of the water layer on silica versus temperature was also analyzed by FTIR spectroscopy and correlated to the water layers morphology. A
qualitative analysis of the spectrums obtained gives two kinds of water layers~: liquid layers, which are removed above $100^\circ$C and strongly
adsorbed water molecules with a solid-like behavior that still remain at $150^\circ$C and under vacuum. In this latter case, the water layers
become more structured due to the proximity of the substrate. For the liquid structure, we assume a molecular mobility that is weakly dependent on
the thickness of the water film. When the temperature increases, the decrease of these forces is related to a structural modification of the water
layers.

\bigskip

The comparison between the evolution of the damping coefficient as a function of time deduced from the Tapping experiments (fig.\ref
{figevoldissip}) and the one of the normalized pull off force (fig.\ref {figevoladh}) as a function of the temperature, suggests that the first
Tapping measurement performed at time zero corresponds to the data measured at $120^\circ$C with the Contact mode (arrow $1$ on both figures).
Thus we can deduce an equivalence between time exposure of the sample to dry conditions and temperature of the sample. The covering of the silica
going on with time, e.g. the quantity of the water molecules adsorbed on the surface increasing, the damping coefficient decrease can be
interpreted as a decrease of the equivalent temperature of the sample (from arrow $1$ to $2$ on fig.\ref {figevoldissip} and from $2$ to $1$ on
fig.\ref{figevoladh}).

\bigskip

{\Large Acknowledgements}

The authors thank the R\'{e}gion Aquitaine for financial support.

\section{Conclusion}

Dynamic Force Microscopy experiments were performed to investigate the
growth of water layers on a silica surface exposed to a rather dry
atmosphere as a function of time. Noncontact dissipation due to the
attractive interaction between the tip and the surface is included in the
theoretical developments. This allows to describe the evolution of the OTCL
phase as a function of time that corresponds to changes of the water layers
structure. The equations obtained from the model are able to reproduce with
a good agreement the experimental observations. These variations can only be
reproduced from the model taking into account rather long relaxation times.
The damping coefficient deduced shows a marked decrease as a function of
time. As a consequence, the evolution of the mechanical susceptibility of
the silica surface indicates the contamination of the surface with water
molecules as a function of time.

These DFM observations are in good agreement with previous tribological and
infra-red measurements for which the contamination of a silica surface with
water molecules was investigated as a function of the sample
temperature.\newpage

\section*{References}

\bibliographystyle{unsrt}

\newpage

\section*{Figures}

\end{document}